\documentclass[prd,amsmath,amssymb,showpacs,superscriptaddress,nofootinbib,twocolumn]{revtex4-1}

\usepackage{graphicx}
\usepackage{epsfig,graphics,subfigure,psfrag,amsmath,amssymb}
\usepackage{lineno}
\usepackage{dcolumn}
\usepackage{bm}
\usepackage{overpic}
\usepackage{xspace}

\usepackage{rotating}
\usepackage{color}
\usepackage{multirow}
\usepackage{colortbl}
\usepackage{epstopdf}
\usepackage[colorlinks,linkcolor=blue,anchorcolor=blue,citecolor=blue]{hyperref}

\newcommand{\jpsi}{J/\psi}
\newcommand{\jp}{J/\psi}
\newcommand{\pio}{\pi^{0}}
\newcommand{\pip}{\pi^{+}}
\newcommand{\pim}{\pi^{-}}
\newcommand{\pippim}{\pi^{+}\pi^{-}}

\newcommand{\kk}{K^{+}K^{-}}
\newcommand{\ks}{K_{S}^{0}}
\newcommand{\ksks}{K_{S}^{0}K_{S}^{0}}
\newcommand{\gammapipi}{\gamma\pi^{+}\pi^{-}}
\newcommand{\pipieta}{\pi^{+}\pi^{-}\eta}

\newcommand{\BESIII}{BES\uppercase\expandafter{\romannumeral3}\xspace}

\begin{document}
\title{\boldmath Observation of the $X(2370)$ and search for the $X(2120)$ in $J/\psi\to\gamma K\bar{K} \eta'$}

\author{M.~Ablikim$^{1}$, M.~N.~Achasov$^{10,e}$, P.~Adlarson$^{59}$, S. ~Ahmed$^{15}$, M.~Albrecht$^{4}$, M.~Alekseev$^{58A,58C}$, A.~Amoroso$^{58A,58C}$, Q.~An$^{55,43}$, ~Anita$^{21}$, Y.~Bai$^{42}$
% ...
\\
      \vspace{0.2cm}
      (BESIII Collaboration)\\
      \vspace{0.2cm} {\it
$^{1}$ Institute of High Energy Physics, Beijing 100049, People's Republic of China\\
% ...
$^{4}$ Bochum Ruhr-University, D-44780 Bochum, Germany\\
% ...
$^{10}$ G.I. Budker Institute of Nuclear Physics SB RAS (BINP), Novosibirsk 630090, Russia\\
% ...
$^{15}$ Helmholtz Institute Mainz, Johann-Joachim-Becher-Weg 45, D-55099 Mainz, Germany\\
% ...
$^{21}$ Indian Institute of Technology Madras, Chennai 600036, India\\
% ...
$^{42}$ Southeast University, Nanjing 211100, People's Republic of China\\
$^{43}$ State Key Laboratory of Particle Detection and Electronics, Beijing 100049, Hefei 230026, People's Republic of China\\
% ...
$^{55}$ University of Science and Technology of China, Hefei 230026, People's Republic of China\\
% ...
$^{58}$ (A)University of Turin, I-10125, Turin, Italy; (B)University of Eastern Piedmont, I-15121, Alessandria, Italy; (C)INFN, I-10125, Turin, Italy\\
$^{59}$ Uppsala University, Box 516, SE-75120 Uppsala, Sweden\\
% ...
$^{e}$ Also at the Novosibirsk State University, Novosibirsk, 630090, Russia\\
% ...
}
}

\date{\today}

\begin{abstract}
Using a sample of $1.31\times10^{9} ~J/\psi$ events collected with the BESIII detector, we perform a study of $J/\psi\to\gamma K\bar{K}\eta'$.
The $X(2370)$ is observed in the $K\bar{K}\eta'$ invariant-mass distribution with a statistical significance of 8.3$\sigma$.
Its resonance parameters are measured  to be
$M=2341.6\pm 6.5\text{(stat.)}\pm5.7\text{(syst.)}$~MeV/$c^{2}$ and $\Gamma = 117\pm10\text{(stat.)}\pm8\text{(syst.)}$~MeV.
The product branching fractions for $\jp\to \gamma X(2370),X(2370)\to \kk\eta'$ and
$\jp\to \gamma X(2370),X(2370)\to \ksks\eta'$ are determined to be
$(1.79\pm0.23\text{(stat.)}\pm0.65\text{(syst.)})\times10^{-5}$ and
$(1.18\pm0.32\text{(stat.)}\pm0.39\text{(syst.)})\times10^{-5}$, respectively.
No evident signal for the $X(2120)$ is observed in the $K\bar{K}\eta'$ invariant-mass distribution.
The upper limits for the product branching fractions of $\mathcal{B}(\jp\to\gamma X(2120)\to\gamma K^{+} K^{-} \eta')$
and $\mathcal{B}(\jp\to\gamma X(2120)\to\gamma K_{S}^{0} K_{S}^{0} \eta')$ are determined to be $1.49\times10^{-5}$ and $6.38\times10^{-6}$ at the 90\% confidence level, respectively.
\end{abstract}

\maketitle

\section{INTRODUCTION}
Quantum chromodynamics (QCD), a non-Abelian gauge field theory, predicts the existence of new types of hadrons with explicit gluonic degrees of freedom (e.g., glueballs, hybrids)~\cite{bibg1,bibg2,bibg3}.
The search for glueballs is an important field of research in hadron physics.
It is, however, challenging since possible mixing of pure glueball states with nearby $q\bar{q}$ nonet mesons
makes the identification of glueballs difficult in both experiment and theory.
Lattice QCD (LQCD) predicts the
lowest-lying glueballs which are scalar (mass 1.5$-$1.7 GeV/$c^2$), tensor (mass 2.3$-$2.4
GeV/$c^2$), and pseudoscalar (mass 2.3$-$2.6 GeV/$c^2$)~\cite{bib3}.  Radiative $J/\psi$ decay is a gluon-rich process and it is therefore regarded as one of the
most promising hunting grounds for glueballs~\cite{bibjpsi1,bibjpsi2}.
Recently, three states, the $X(1835)$, $X(2120)$ and $X(2370)$, are observed in the BESIII experiment in the $\pi^{+}\pi^{-}\eta'$ invariant-mass distribution through the decay of $J/\psi\to\gamma\pi^{+}\pi^{-}\eta'$ with statistical significances larger than 20$\sigma$, 7.2$\sigma$ and 6.4$\sigma$, respectively~\cite{PRL1}.
The measured mass of the $X(2370)$ is consistent with the pseudoscalar glueball candidate predicted by LQCD calculations~\cite{bib3}.
 In the case of a pseudoscalar glueball, the branching fractions of the $X(2370)$ decaying into $KK\eta'$ and $\pi\pi\eta'$
are predicted to be 0.011 and 0.090~\cite{PRD1}, respectively, in accordance with calculations that are based upon the chiral effective Lagrangian.
Study on the decays to $K\bar{K}\eta'$ of the glue-ball candidate X states is helpful to identify their natures.

In this paper,
the $X(2370)$ as well as the $X(2120)$ are studied via the decays of $J/\psi\to\gamma K^{+}K^{-}\eta'$ and
$J/\psi\to\gamma K_{S}^{0}K_{S}^{0}\eta'$($K_{S}^{0}\to\pi^{+}\pi^{-}$) using (1310.6$\pm$7.0)$\times$10$^6$ $J/\psi$ decays~\cite{jpsinumber} collected with the BESIII detector in 2009 and 2012.
Two $\eta'$ decay modes are used, namely
$\eta'\to\gamma\rho^{0}(\rho^{0}\to\pi^{+}\pi^{-})$ and $\eta'\to\pi^{+}\pi^{-}\eta(\eta\to\gamma\gamma)$.

\section{DETECTOR AND MONTE CARLO SIMULATIONS}
The BESIII detector is a magnetic
spectrometer~\cite{Ablikim:2009aa} located at the Beijing Electron
Positron Collider II(BEPCII)~\cite{Yu:IPAC2016-TUYA01}. The
cylindrical core of the BESIII detector consists of a helium-based
 multilayer drift chamber (MDC), a plastic scintillator time-of-flight
system (TOF), and a CsI(Tl) electromagnetic calorimeter (EMC),
which are all enclosed in a superconducting solenoidal magnet
providing a 1.0~T (0.9~T in
2012) magnetic field. The solenoid is supported by an
octagonal flux-return yoke with resistive plate counter muon
identifier modules interleaved with steel. The acceptance of
charged particles and photons is 93\% over $4\pi$ solid angle. The
charged-particle momentum resolution at $1~{\rm GeV}/c$ is
$0.5\%$, and the $dE/dx$ resolution is $6\%$ for the electrons
from Bhabha scattering. The EMC measures photon energies with a
resolution of $2.5\%$ ($5\%$) at $1$~GeV in the barrel (end cap)
region. The time resolution of the TOF barrel part is 68~ps, while
that of the end cap part is 110~ps.

Simulated samples produced with the {\sc
geant4}-based~\cite{geant4} Monte Carlo (MC) package which
includes the geometric description of the BESIII detector and the
detector response, are used to determine the detection efficiency
and to estimate the backgrounds. The simulation includes the beam
energy spread and initial-state radiation (ISR) in the $e^+e^-$
annihilations modeled with the generator {\sc
kkmc}~\cite{ref:kkmc}. The inclusive MC sample consists of the production of the $J/\psi$
resonance, and the continuum processes incorporated in {\sc
kkmc}~\cite{ref:kkmc}. The known decay modes are modeled with {\sc
evtgen}~\cite{ref:evtgen} using branching fractions taken from the
Particle Data Group~\cite{pdg}, and the remaining unknown decays
from the charmonium states are generated with {\sc
lundcharm}~\cite{ref:lundcharm}. The final-state radiations (FSR)
from charged final-state particles are incorporated with the {\sc
photos} package~\cite{photos}.
Background is studied
using a sample of $1.2 $$\times$$ 10^{9}$ simulated $\jp$ events.
Phase-space (PHSP) MC samples of $\jpsi\to\gamma \kk\eta'$ and $\jpsi\to\gamma \ksks\eta'$ are generated to describe the nonresonant contribution.
To estimate the selection efficiency and to optimize the selection criteria,
signal MC events are generated for $\jp\to\gamma X(2120)/X(2370)
\to\gamma\kk\eta'$ and $\jp\to\gamma X(2120)/X(2370)\to\gamma\ksks\eta'$
channel, respectively. The polar angle of the photon in the $\jp$ center-of-mass system, $\theta_{\gamma}$, follows a  $1+ \mathrm{cos}^{2}\theta_{\gamma}$ function.
For the process of $\eta'\to\gamma\rho^{0}, \rho^{0}\to\pi^{+}\pi^{-}$,
a  generator taking into account both the $\rho - \omega$ interference and the box
anomaly is used~\cite{gammapipiDIY}.
The analysis is performed in the framework of the BESIII offline software system
(BOSS)~\cite{ref:boss} incorporating
the detector calibration, event reconstruction and data storage.

\section{EVENT SELECTION}
Charged-particle tracks in the polar angle range $|\cos\theta| < 0.93$ are reconstructed from hits in the MDC.
Tracks (excluding those from $K_{S}^{0}$ decays) are selected that extrapolated to be within 10 cm
from the interaction point in the beam direction and 1 cm in the plane perpendicular to the beam.
The combined information from energy-loss ($dE/dx$) measurements in the MDC and time
in the TOF is used to obtain confidence levels for particle identification (PID) for $\pi$, $K$ and $p$ hypotheses.
For $\jpsi\to\gamma\kk\eta'$ decay,
each track is assigned to the particle type corresponding to the highest confidence level;
candidate events are required to have four charged tracks with zero net charge and with two
opposite charged tracks identified as kaons and the other two identified as pions.
For the $\jpsi\to\gamma\ksks\eta'$ decay, each track is assumed to be a pion and no PID restrictions are applied;
candidate events are required to have six charged tracks with zero net charge.
$\ks$ candidates are reconstructed from a secondary vertex fit to all $\pippim$ pairs,
and each $\ks$ candidate is required to satisfy $|M_{\pip\pim}-m_{\ks}|<$9~MeV/$c^{2}$, where $m_{\ks}$ is the nominal mass of the $\ks$~\cite{pdg}. The reconstructed $\ks$ candidates are used as an input for the subsequent kinematic fit.

Photon candidates are required to have an energy deposition above 25 MeV in the barrel region ($|\cos\theta|<0.80$) and 50 MeV in the end cap ($0.86<|\cos\theta|<0.92$). To exclude showers from charged tracks, the angle between the shower position and the charged tracks extrapolated to the EMC must be greater than $5^{\circ}$. A timing requirement in the EMC is used to suppress electronic noise and energy deposits unrelated to the event. At least two (three) photons are required for the $\eta'\to\gamma\rho^{0}$ ($\eta'\to\pipieta$) mode.

\begin{figure}[htbp]
\centering
\includegraphics[width=0.24\textwidth]{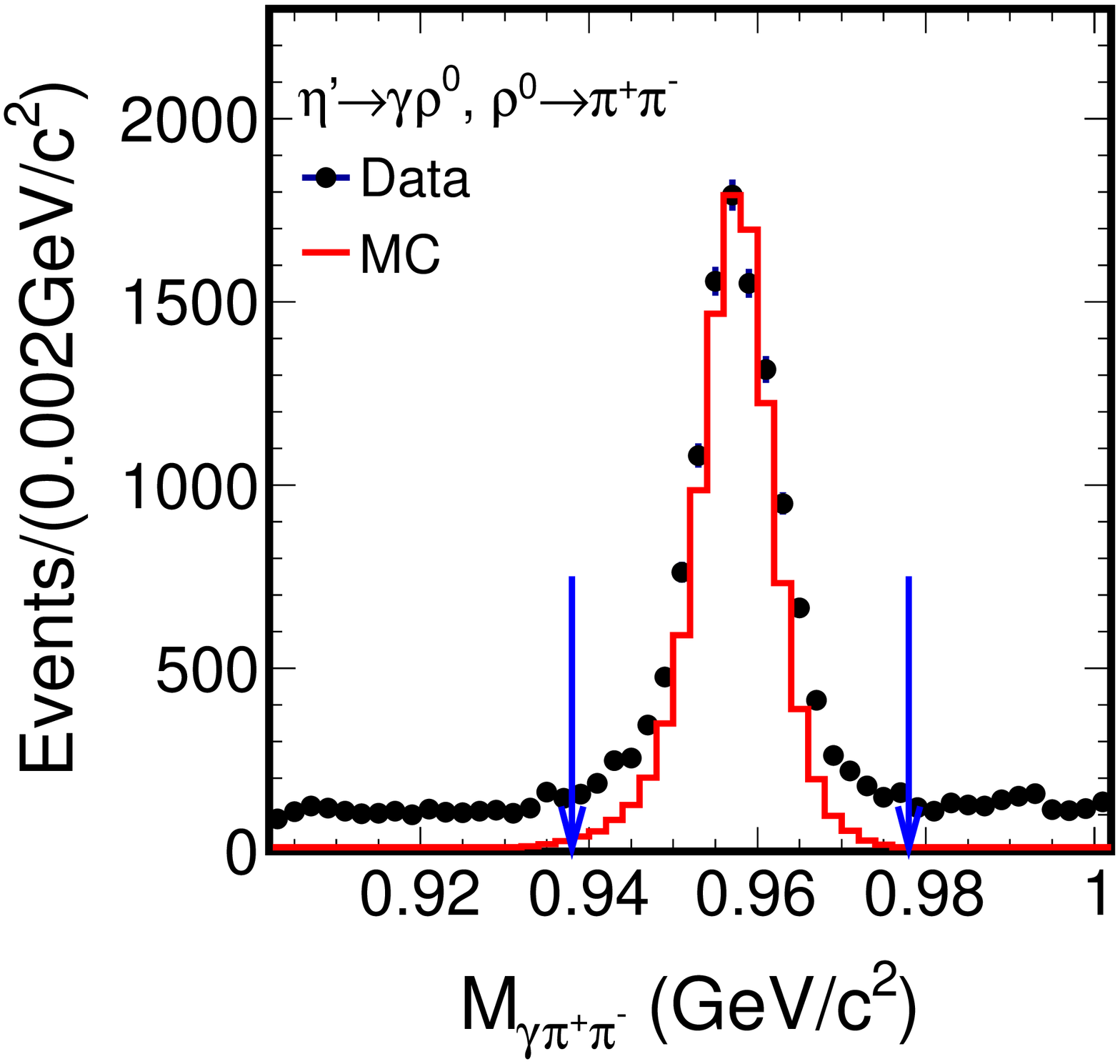}\put(-30,85){{(a)}}
\includegraphics[width=0.24\textwidth]{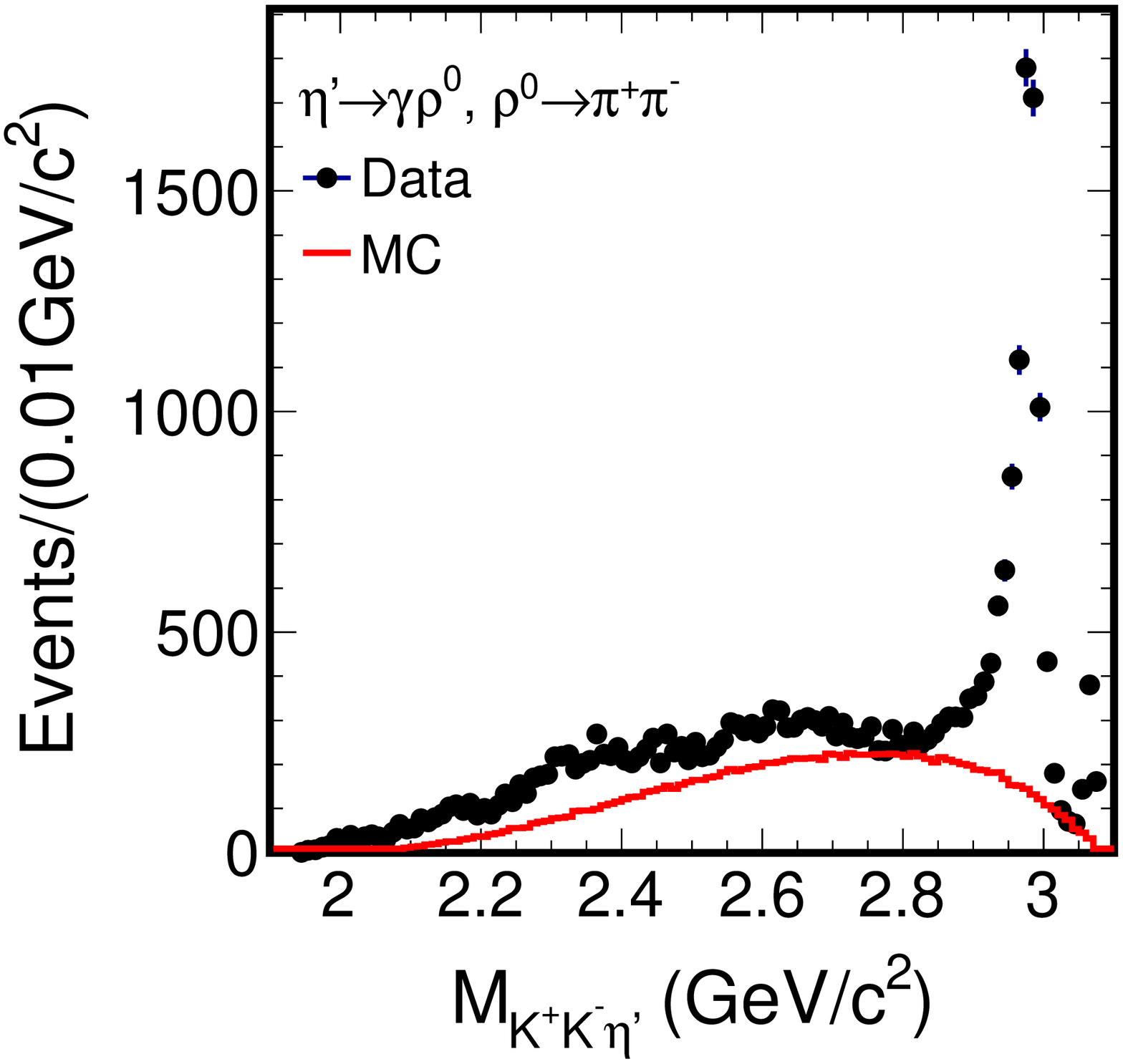}\put(-30,85){{(b)}}
\vskip -0.03cm
\includegraphics[width=0.24\textwidth]{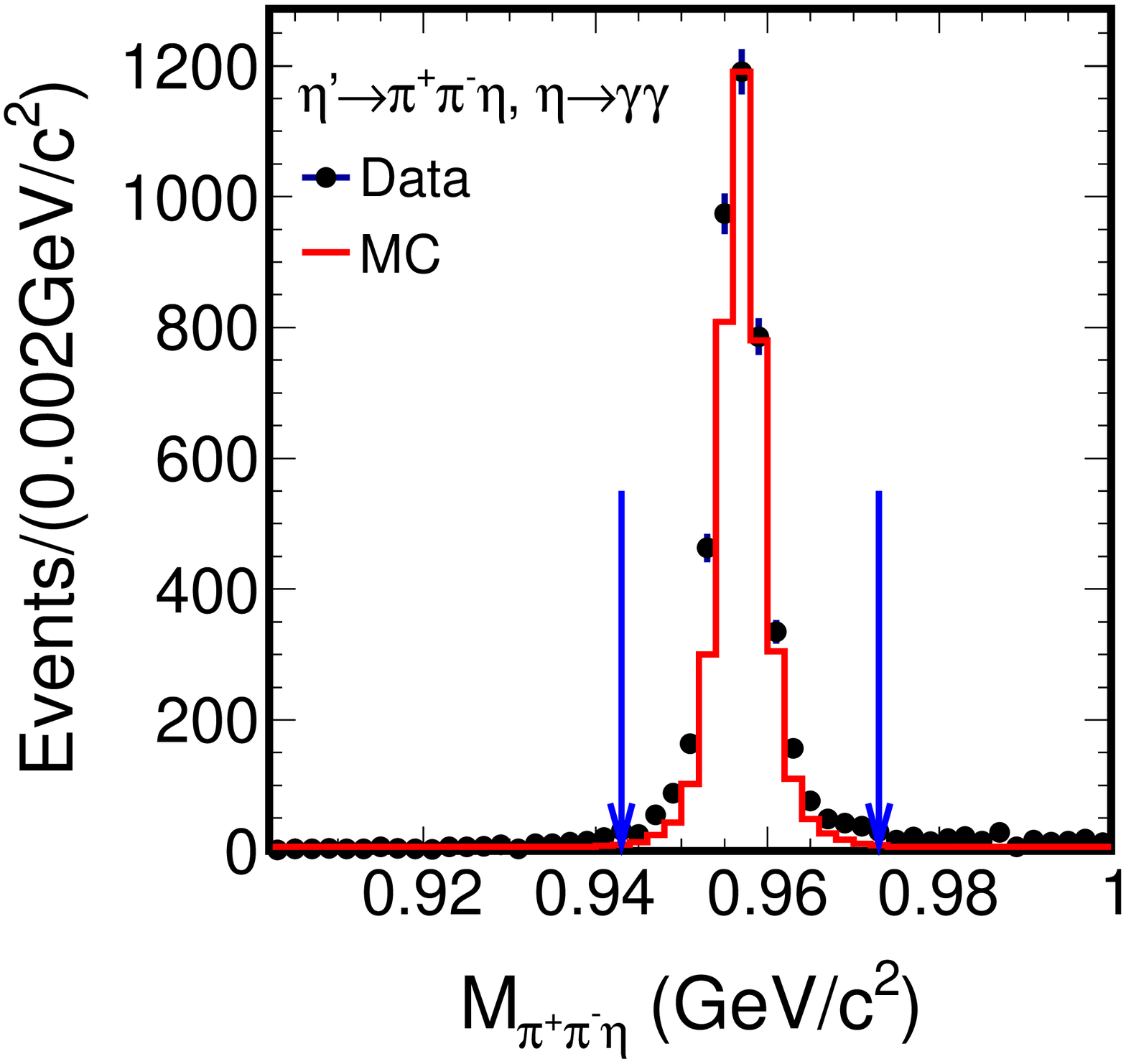}\put(-30,85){{(c)}}
\includegraphics[width=0.24\textwidth]{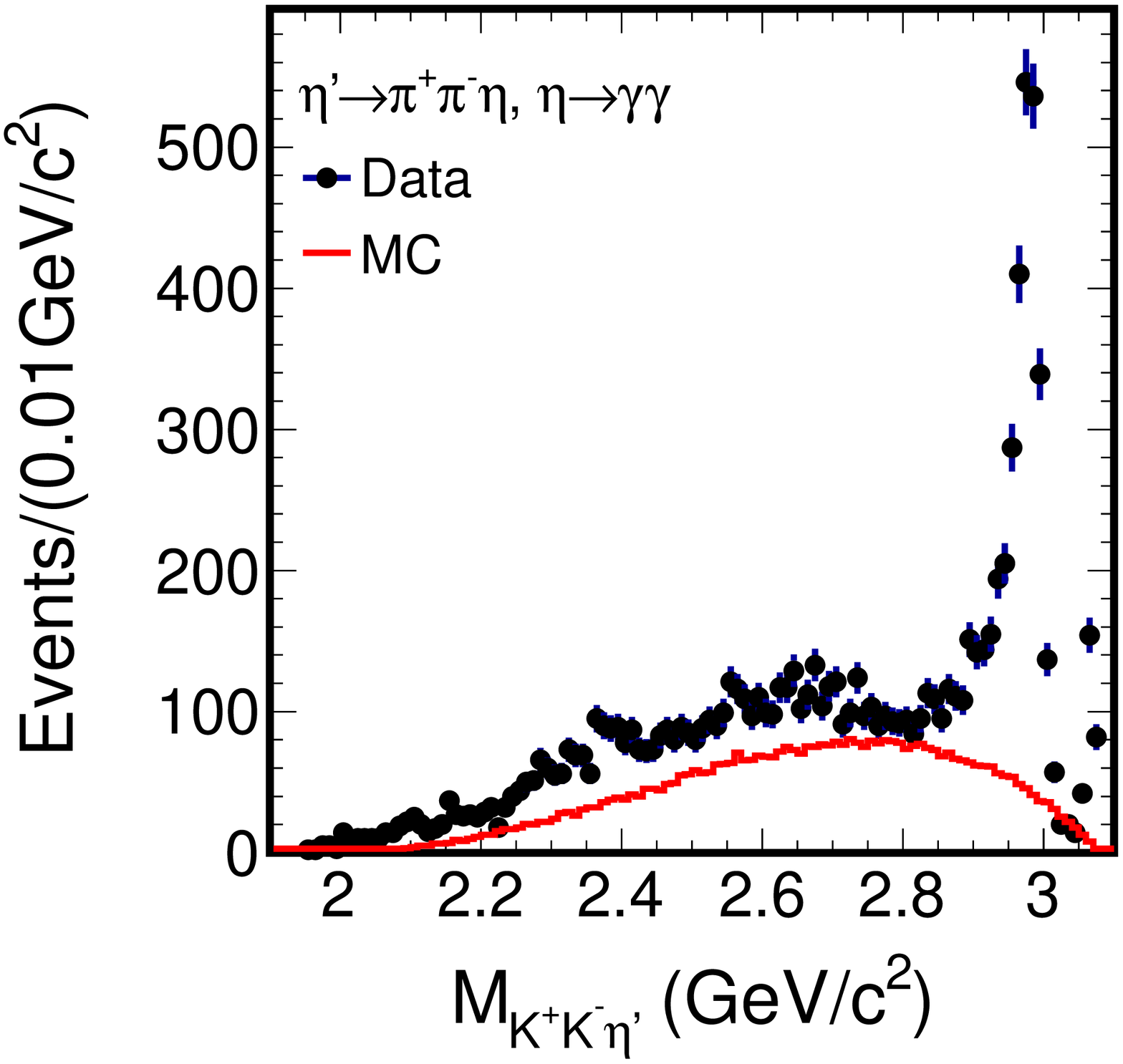}\put(-30,85){{(d)}}
\caption{Invariant-mass distributions for the selected candidates of $\jpsi\to\gamma\kk\eta'$.
     Plots (a) and (b) are invariant-mass distributions of $\gammapipi$
     and $\kk\eta'$ for $\eta'\to\gamma\rho^{0}$, $\rho^{0}\to\pi^{+}\pi^{-}$, respectively;
     plots (c) and (d) are the invariant-mass distributions of $\pipieta$
     and $\kk\eta'$ for $\eta'\to\pipieta$, $\eta\to\gamma\gamma$, respectively.
      The dots with error bars correspond to data and the histograms are the results of PHSP MC simulations (arbitrary normalization).}

\label{selectkketap}
\end{figure}

For the $\jpsi\to\gamma\kk\eta' (\eta'\to\gamma\rho^{0})$ channel, a four-constraint (4C) kinematic fit is performed by requiring the total energy and each momentum component to be conserved to the hypothesis of
$\jpsi\to\gamma\gamma\kk\pip\pim$. For events with more than two photon candidates, the combination with the minimum $\chi_{4C}^{2}$ is selected, and $\chi_{4C}^{2}<$ 25 is required.
Events with $|M_{\gamma\gamma} - m_{\pi^{0}}| <$ 30~MeV/$c^{2}$ or $|M_{\gamma\gamma} - m_{\eta}| <$~30~MeV/$c^{2}$ are rejected to suppress background containing  $\pi^{0}$ or $\eta$, where the $m_{\pi^{0}}$ and $m_{\eta}$ are the nominal masses of $\pi^{0}$ and $\eta$~\cite{pdg}.
A clear $\eta'$ signal is observed in the invariant-mass distribution of $\gammapipi$ ($M_{\gamma\pi^{+}\pi^{-}}$),
as shown in Fig.~\ref{selectkketap}(a).
Candidates of $\rho$ and $\eta'$ are reconstructed from the $\pippim$ and $\gammapipi$ combinations with 0.55~GeV/$c^{2} < M_{\pi^{+}\pi^{-}} < $ 0.85~GeV/$c^{2}$ and $|M_{\gamma\pi^{+}\pi^{-}} - m_{\eta'}| <$ 20~MeV/$c^{2}$,
where $m_{\eta'}$ is the nominal mass of $\eta'$~\cite{pdg}, respectively.
If there are more than one combination satisfing the selection criteria, the combination with $M_{\gamma\pi^{+}\pi^{-}}$ closest to $m_{\eta'}$ is selected.
After applying the above requirements, we obtain the invariant-mass distribution of $\kk\eta'$ ($M_{\kk\eta'}$) as shown in Fig.~\ref{selectkketap}(b).

To reduce background and to improve the mass resolution of the $\jpsi\to\gamma\kk\eta' (\eta'\to\pipieta)$ channel,  a five-constraint (5C) kinematic fit is performed whereby the total four momentum of the final-state particles are constrained to the total initial four momentum of the colliding beams and the invariant mass of the two photons from the decay of the $\eta$ is constrained to its nominal mass.
If there are more than three photon candidates, the combination
with the minimum $\chi_{5C}^{2}$ is retained, and $\chi_{5C}^{2} < $ 45 is required. To suppress background from $\pi^{0}\to\gamma\gamma$,
$|M_{\gamma\gamma} - m_{\pi^{0}}| >$ 30~MeV/$c^{2}$ is required for all photon pairs.
The $\eta'$ candidates are formed from the $\pipieta$ combination satisfying
$|M_{\pi^{+}\pi^{-}\eta} - m_{\eta'}| <$ 15 MeV/$c^{2}$, where $M_{\pi^{+}\pi^{-}\eta}$ is the invariant mass of $\pi^{+}\pi^{-}\eta$, as shown in Fig.~\ref{selectkketap}(c).
After applying the mass restrictions, we obtain the invariant-mass distribution of $\kk\eta'$($\eta'\to\pipieta$) as shown in Fig.~\ref{selectkketap}(d).

\begin{figure}[htbp]
\centering
\includegraphics[width=0.24\textwidth]{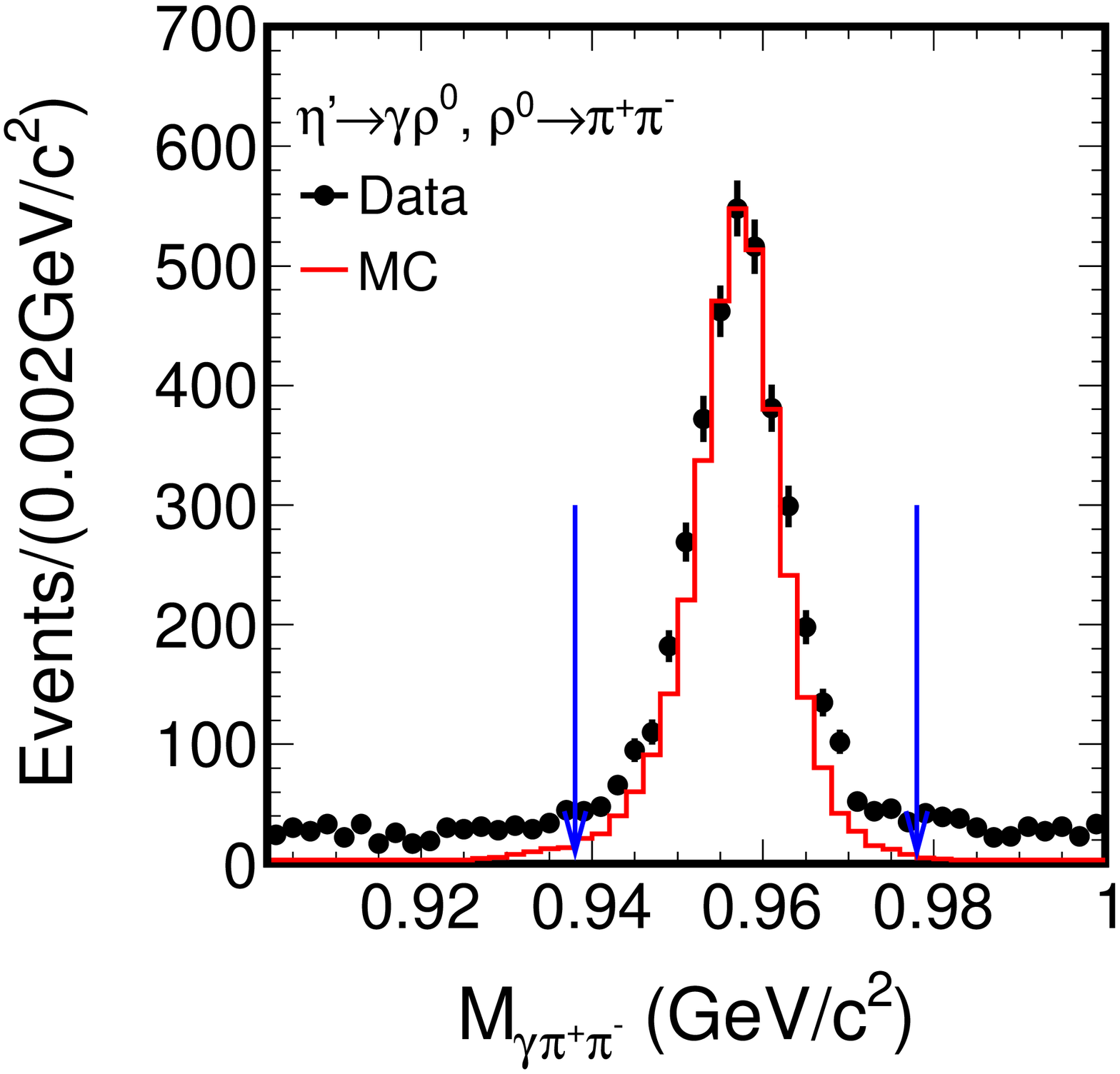}\put(-30,85){{(a)}}
\includegraphics[width=0.24\textwidth]{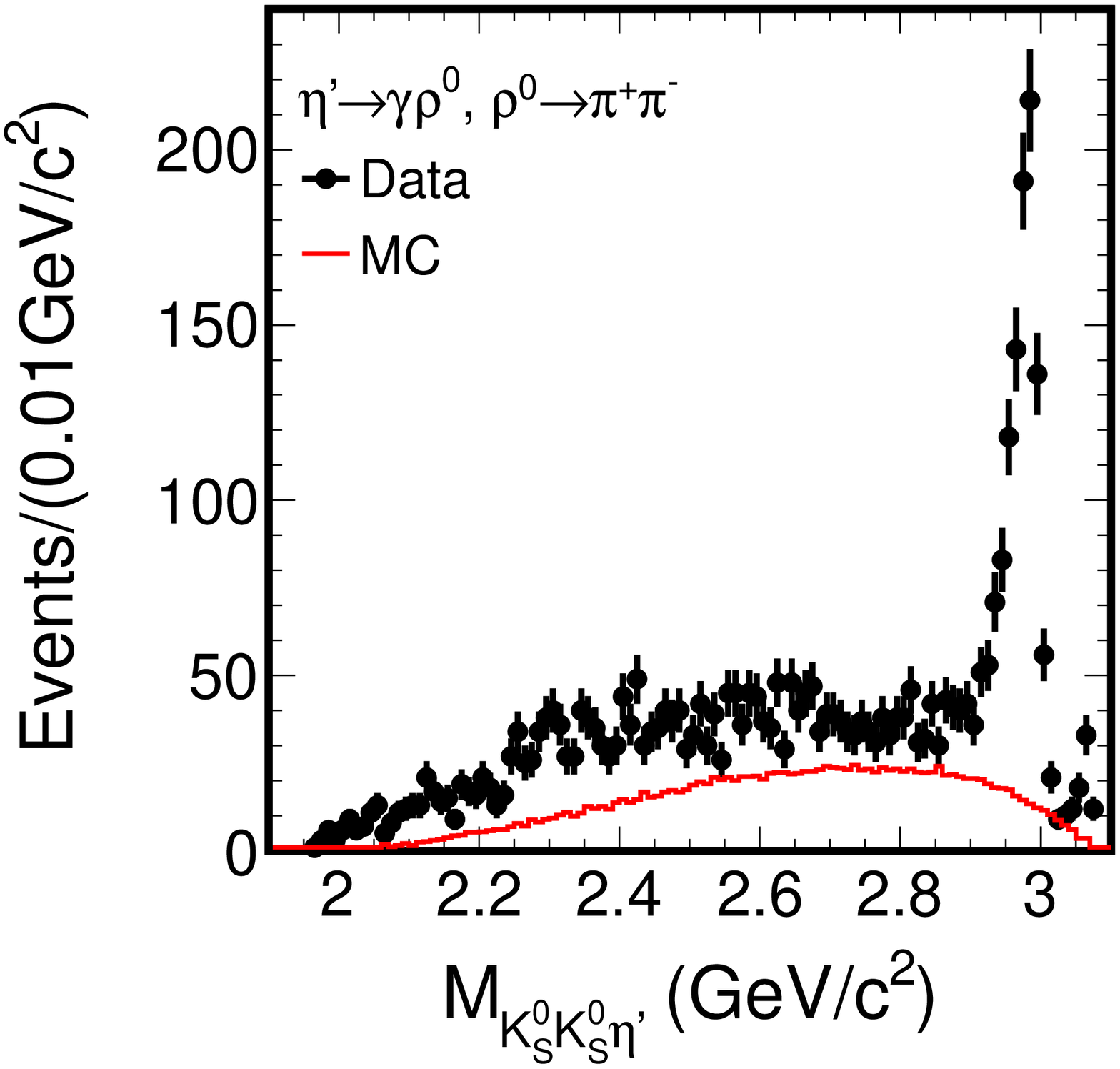}\put(-30,85){{(b)}}
\vskip -0.03cm
\includegraphics[width=0.24\textwidth]{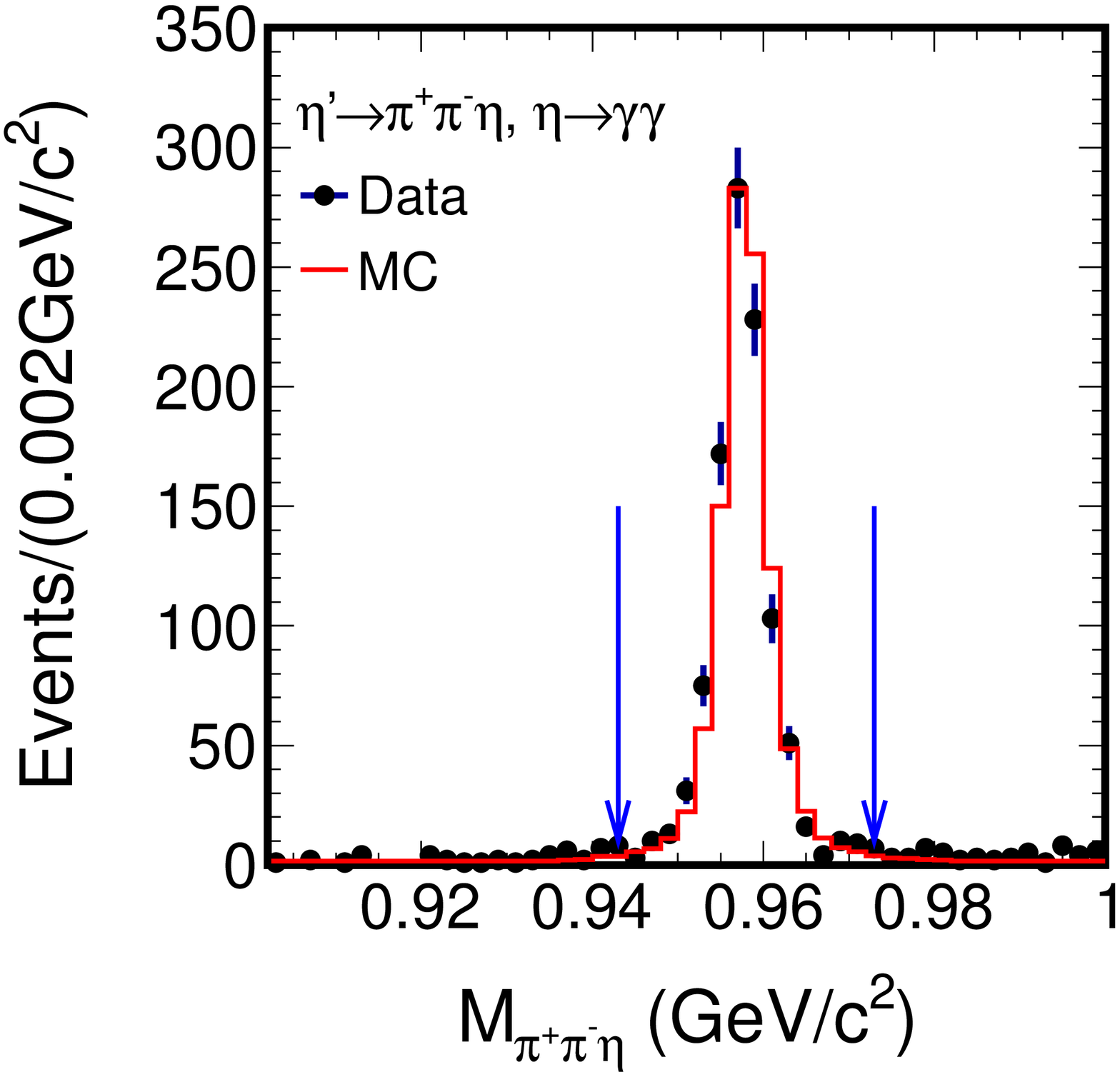}\put(-30,85){{(c)}}
\includegraphics[width=0.24\textwidth]{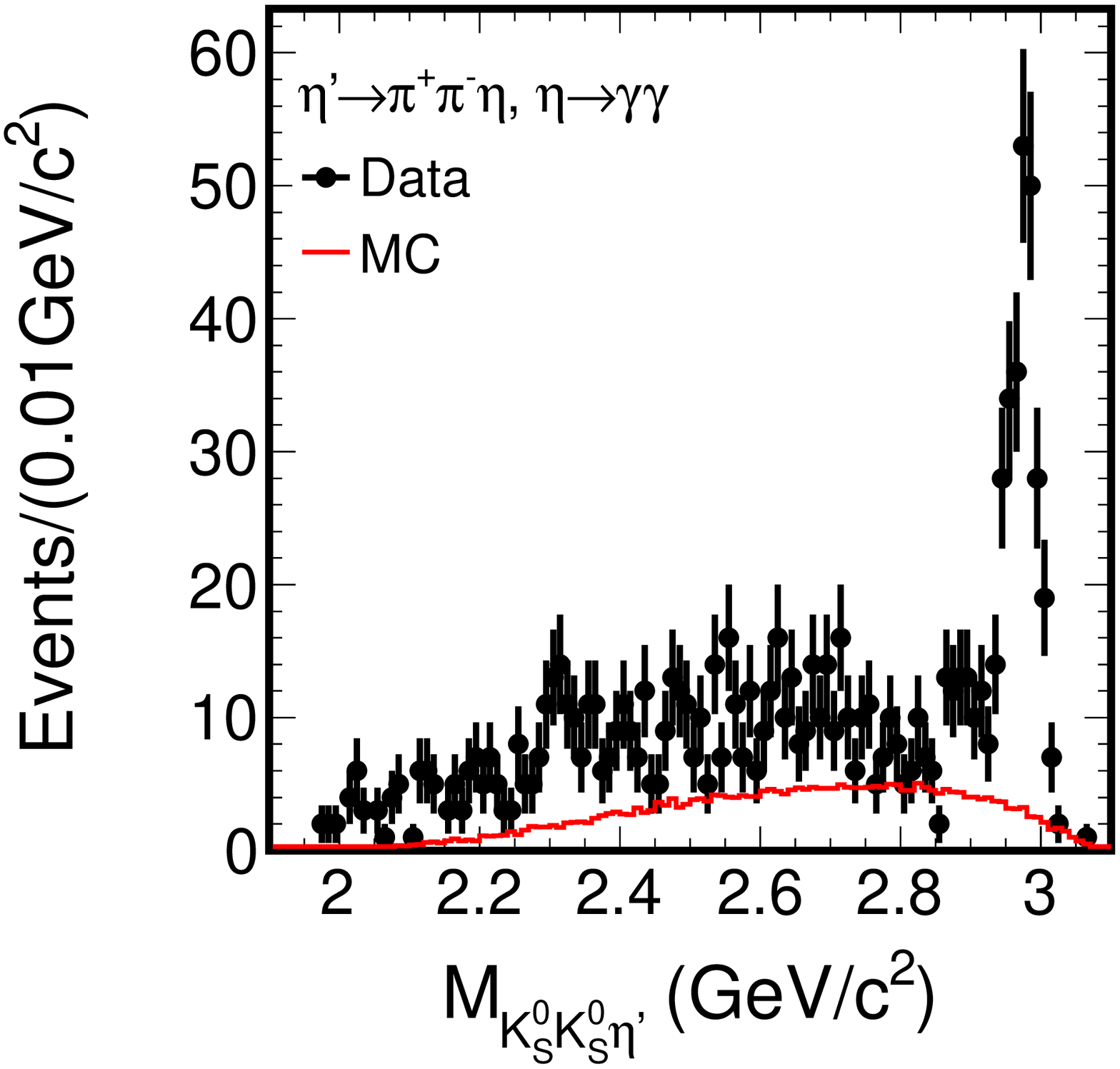}\put(-30,85){{(d)}}
\caption{Invariant-mass distributions for the selected $\jpsi\to\gamma\ksks\eta'$ candidate events.
     (a) and (b) are the  invariant-mass distributions of $\gammapipi$
     and $\ksks\eta'$ for $\eta'\to\gamma\rho^{0}$, $\rho^{0}\to\pi^{+}\pi^{-}$, respectively;
     (c) and (d) are the  invariant-mass distribution of $\pipieta$
     and $\ksks\eta'$ for $\eta'\to\pipieta$, $\eta\to\gamma\gamma$, respectively.
           The dots with error bars represent the data and the histograms are the results of  PHSP MC simulations (arbitrary normalization).}

\label{selectksksetap}
\end{figure}

\begin{figure*}[htbp]
\centering
\includegraphics[width=0.45\textwidth]{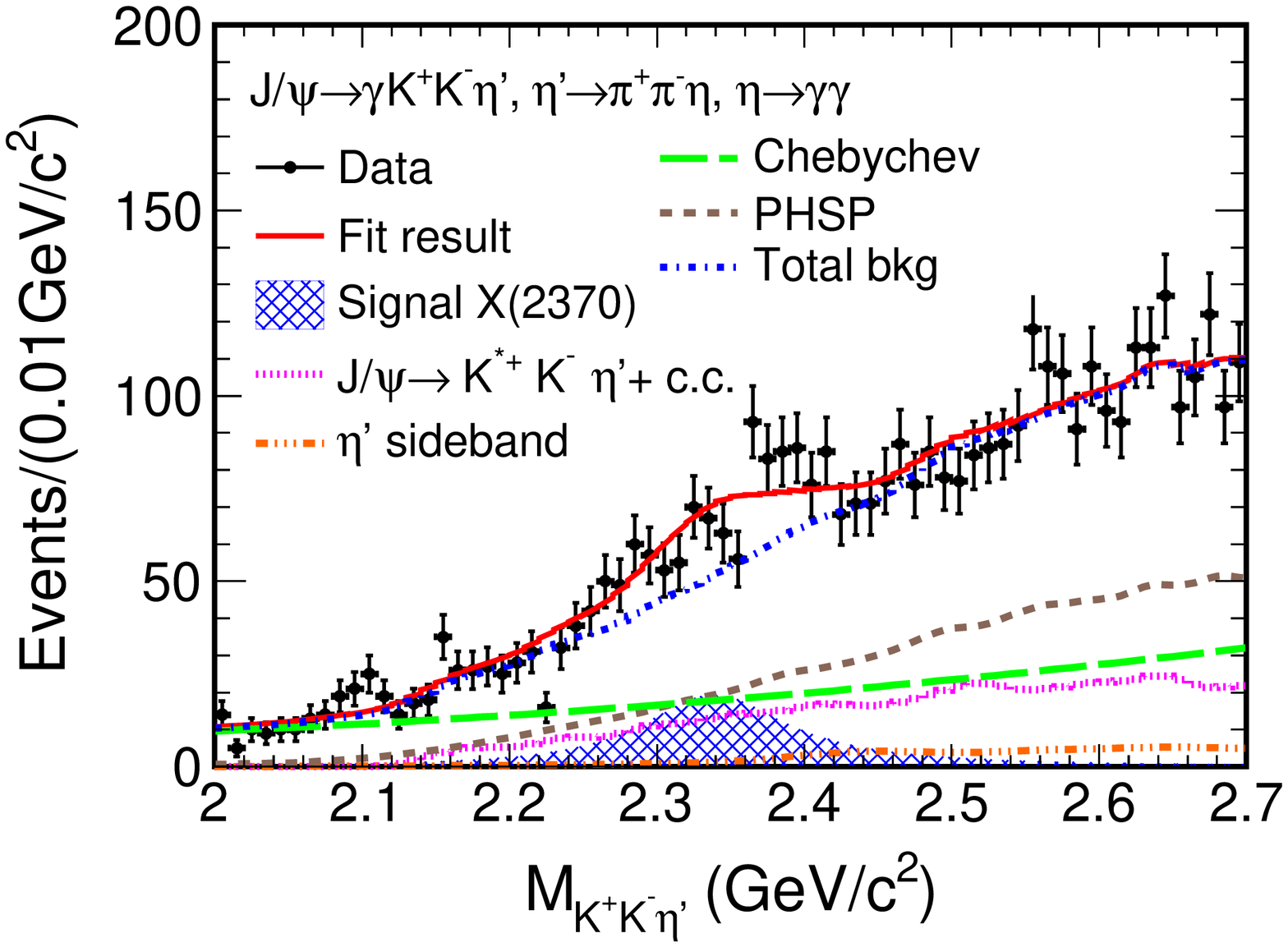}\put(-30,145){{(a)}}
\includegraphics[width=0.45\textwidth]{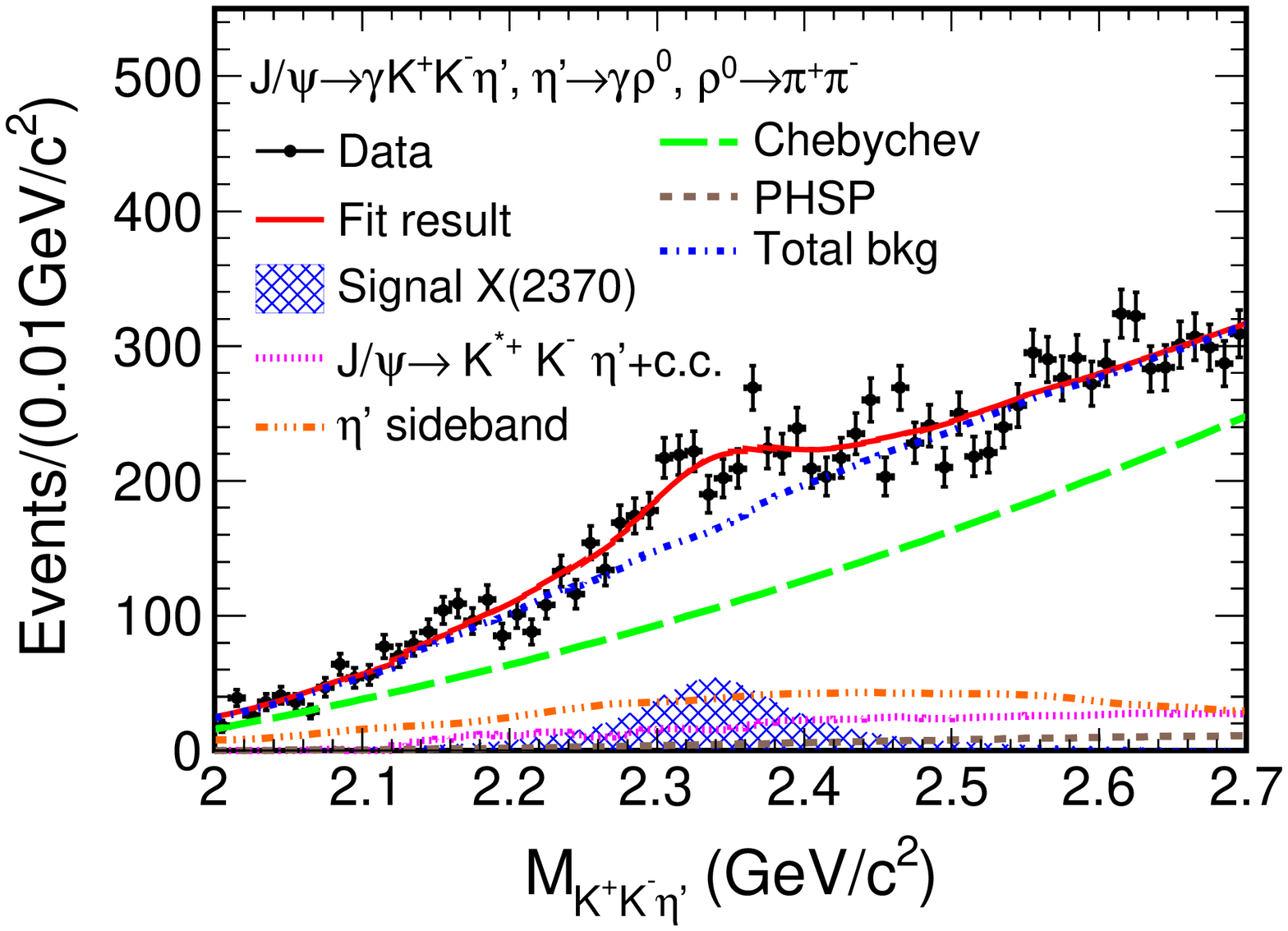}\put(-30,145){{(b)}}
\vskip -0.03cm
\includegraphics[width=0.45\textwidth]{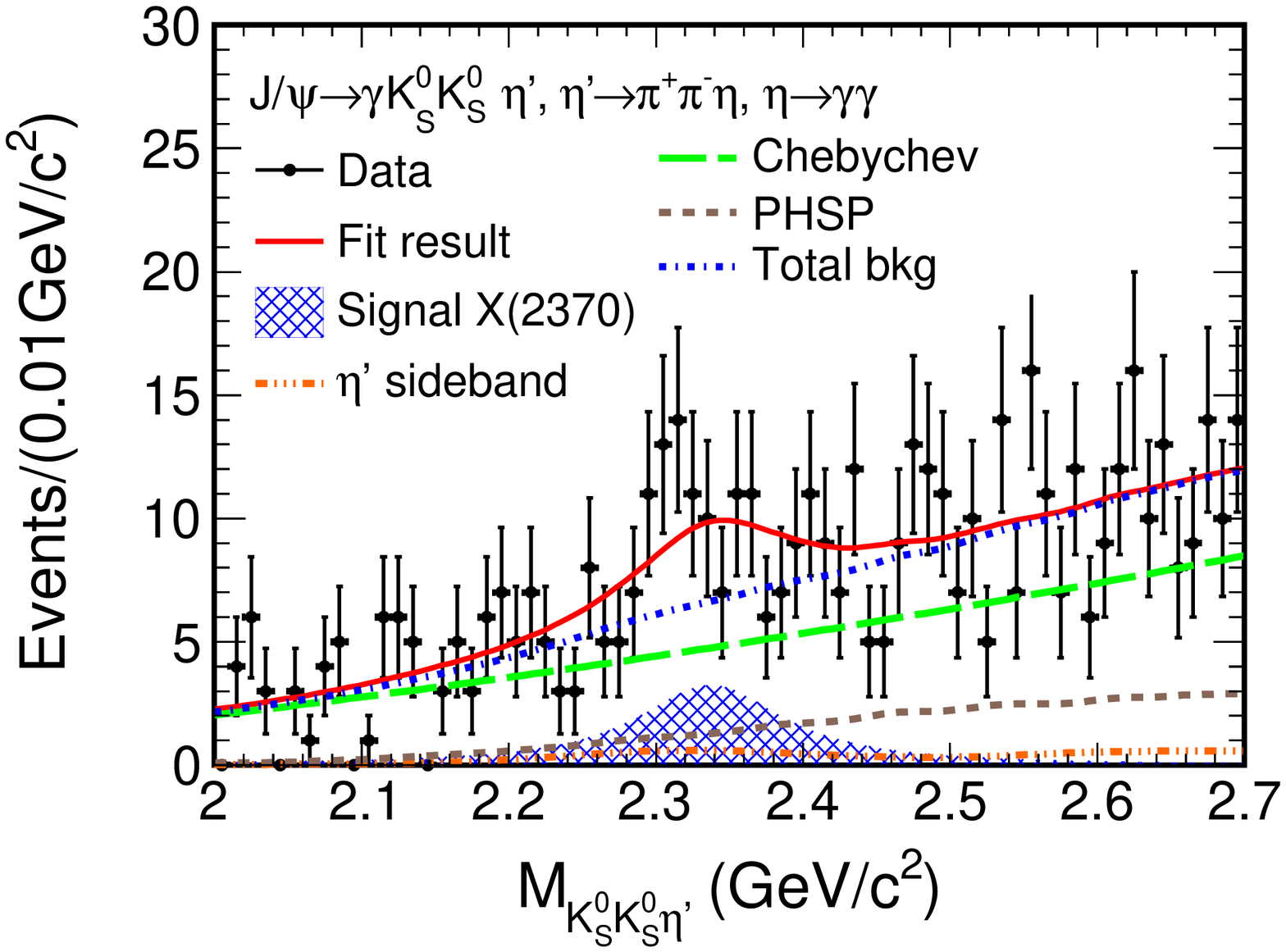}\put(-30,145){{(c)}}
\includegraphics[width=0.45\textwidth]{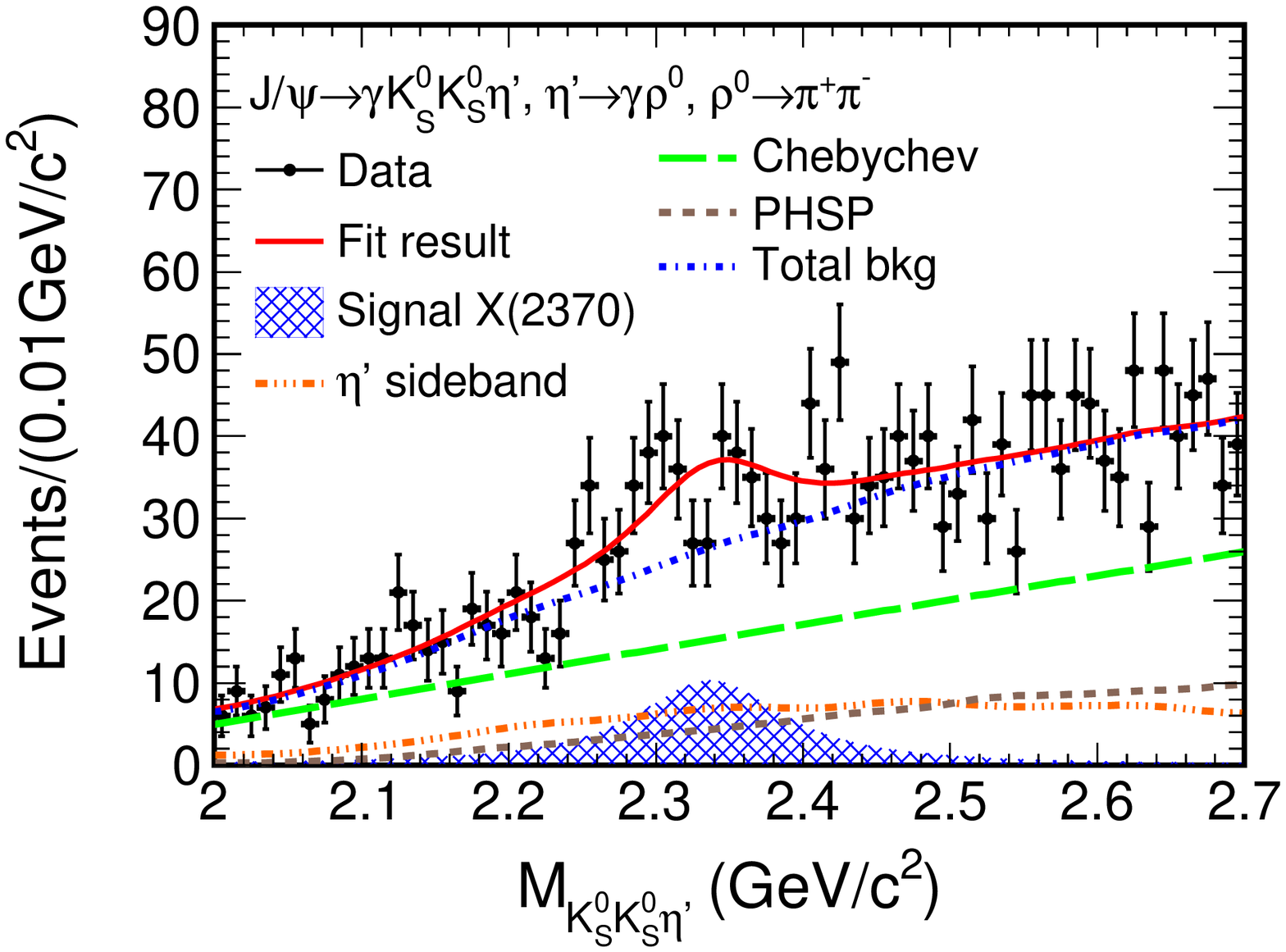}\put(-30,145){{(d)}}
\caption{The fit result for  $X(2370)$ in the invariant-mass distribution of $K \bar K \eta'$
     for the decays: (a) $\jp \to \gamma X(2370),X(2370)\to \gamma\kk\eta', \eta'\to\pipieta$, $\eta\to\gamma\gamma$,
     (b)  $\jp \to \gamma X(2370),X(2370)\to \gamma\kk\eta', \eta'\to\gamma\rho^{0}, \rho^{0}\to\pi^{+}\pi^{-}$,
     (c)   $\jp \to \gamma X(2370),X(2370)\to \gamma\ksks\eta', \eta'\to\pipieta$, $\eta\to\gamma\gamma$,
     and (d)  $\jp \to \gamma X(2370),X(2370)\to \gamma\ksks\eta', \eta'\to\gamma\rho^{0}, \rho^{0}\to\pi^{+}\pi^{-}$.
     The dots with error bars represent the data;
     the solid curves show the fit results;
     the grid areas represent the signal of the $X(2370)$;
     the dotted lines are the background shapes from $J/\psi\to K^{*+}K^{-}\eta'+c.c.$;
     the short dashed double dotted lines show the $\eta'$ sidebands;
     the long dashed lines represent the Chebychev polynomial function;
     the grey short dashed lines are the contribution from  PHSP MC
          and the dashed dotted lines show the sum of all backgrounds.}

\label{fit2370}
\end{figure*}

For the $\jpsi\to\gamma\ksks\eta' (\eta'\to\gamma\rho^{0})$ channel,
the $\gamma \gamma \ksks \pi^+\pi^-$ candidates are subjected to a 4C kinematic fit.
For events with more than two photons or two $K^{0}_{S}$ candidates, the combination with the smallest $\chi^{2}_{4C}$ is retained, and  $\chi^{2}_{4C} < 45$ is required.
To suppress background events containing a $\pi^{0}$ or $\eta$, events with $|M_{\gamma\gamma} - m_{\pi^{0}}| <$ 30~MeV/$c^{2}$ or $|M_{\gamma\gamma} - m_{\eta}| <$ 30~GeV/$c^{2}$ are rejected.
The $\pippim$ invariant mass is required to be in the $\rho$ mass region, 0.55~GeV/$c^{2} < M_{\pi^{+}\pi^{-}} < $ 0.85~GeV/$c^{2}$, and
$|M_{\gamma\pi^{+}\pi^{-}} - m_{\eta'}| <$ 20~MeV/$c^{2}$ is applied to select $\eta'$ signal.
If more than one combination of $\gamma\pi^{+}\pi^{-}$ are obtained,
the combination with $M_{\gamma\pi^{+}\pi^{-}}$ closest to $m_{\eta'}$ is selected as shown in Fig.~\ref{selectksksetap}(a).
After applying the above requirements, we obtain the $\ksks\eta'$($\eta'\to\gamma\rho^{0}$) invariant-mass spectrum as illustrated in Fig.~\ref{selectksksetap}(b).

Candidate events of the $\jpsi\to\gamma\ksks\eta'$ $(\eta'\to\pipieta)$ channel are subjected to a
5C kinematic fit, which is similar to that for the $\jpsi\to\gamma\kk\eta'$ $(\eta'\to\pipieta)$ mode.
If there are more than three photons or more than two $K^{0}_{S}$ candidates, only the combination with the minimum $\chi_{5C}^{2}$ is selected and $\chi_{5C}^{2} < $50 is required.
To reduce the combinatorial background from $\pi^{0}\to\gamma\gamma$ events, $|M_{\gamma\gamma} - m_{\pi^{0}}| >$ 30~MeV/$c^{2}$ is required for all photon pairs.
For selecting the $\eta'$ signal, the $\pipieta$ combination satisfying
$|M_{\pi^{+}\pi^{-}\eta} - m_{\eta'}| <$ 15~MeV/$c^{2}$ is required, as shown in Fig.~\ref{selectksksetap}(c).
After applying the above selection criteria, we obtain the invariant-mass distribution of $\ksks\eta'$$(\eta'\to\pipieta)$ events as shown in Fig.~\ref{selectksksetap}(d).

\section{SIGNAL EXTRACTION}
Potential backgrounds are studied using an inclusive MC sample of $1.2\times10^{9}$ $\jpsi$ decays.
No significant peaking background is identified in the invariant-mass distributions of
$\kk\eta'$ and $\ksks\eta'$.
Non-$\eta'$ processes are studied using the $\eta'$ mass sidebands.
The major background in the decay $\jpsi\to\gamma\kk\eta'$ stem from $\jpsi\to K^{*+} K^{-}\eta' (K^{*+}\to K^{+}\pi^{0}) + c.c.$. The contribution of $\jpsi\to K^{*+} K^{-}\eta' (K^{*+}\to K^{+}\pi^{0}) + c.c.$ is estimated by the background-subtracted $\kk\eta'$ spectrum of $\jpsi\to K^{*+} K^{-}\eta' (K^{*+}\to K^{+}\pi^{0}) + c.c.$ events selected from data. The spectrum is reweighted according to the ratio of efficiency of $\jpsi\to\gamma\kk\eta'$ and $\jpsi\to K^{*+} K^{-}\eta' (K^{*+}\to K^{+}\pi^{0}) + c.c.$.
For the $\jpsi\to\gamma\ksks\eta'$ case, backgrounds from the process $\jpsi\to\pio\ksks\eta'$ are negligible, as it is forbidden due to charge conjugation invariance.

A structure near 2.34 GeV/$c^{2}$ is observed in the invariant-mass distribution of $\kk\eta'$ and $\ksks\eta'$.
We performed a simultaneous unbinned maximum-likelihood fit to the $\kk\eta'$ and $\ksks\eta'$ invariant-mass distributions between 2.0 and 2.7 GeV/$c^{2}$, as shown in Fig~\ref{fit2370}.
The signal is represented by an efficiency-weighted non-relativistic Breit-Wigner (BW) function convolved with a double Gaussian function to account for the mass resolution.
The mass and width of BW function are left free in the fit while the parameters of the double Gaussian function are fixed on the results obtained from the fit of signal MC  samples generated with zero width.
The non-$\eta'$ background events are described with $\eta'$ sideband data and the yields from these sources are fixed;
the $\jpsi\to K^{*+}K^{-}\eta' + c.c.$ contributions in $\jpsi\to\gamma\kk\eta'$ decay channel are studied as discussed above and its shape as well as the yields are fixed in the fit;
the contribution from the nonresonant $\gamma K\bar{K}\eta'$ production is described by the shape from the
PHSP MC sample of $J/\psi\to\gamma K\bar{K} \eta'$ and its absolute yield is set as a free
parameter in the fit;
the remaining background is described by a second order Chebychev polynomial function and its parameters are left to be free.
In the simultaneous fit, the resonance parameters are free parameters and constrained to be the same for all four channels.
The signal ratio for the two $\eta'$ decay modes is fixed with a factor
calculated by their branching fractions and efficiencies.
The signal ratio between $\jpsi\to\gamma X(2370)\to\gamma\kk\eta'$ and $\jpsi\to\gamma X(2370)\to\gamma\ksks\eta'$ is a free parameter in the fit.
 The obtained mass, width and the number of signal events for the $X(2370)$ are listed in Table~\ref{fitresult2370}.
 A variety of fits with different fit ranges, $\eta'$ sideband regions and background shapes are performed, and the smallest statistical significance among these fits is found to be 8.3$\sigma$.
With the detection efficiencies listed in Table~\ref{effi_all},
the product branching fractions for $\jp\to \gamma X(2370),X(2370)\to \kk\eta'$ and $\jp\to \gamma X(2370),X(2370)\to \ksks\eta'$
are determined to be $(1.79\pm0.23)\times10^{-5}$ and $(1.18\pm0.32)\times10^{-5}$, respectively, where the uncertainties are statistical only.

\begin{table}[htp]
\begin{center}
 \caption{\label{fitresult}Fit results for the structure around 2.34 GeV/$c^{2}$ and 2.12 GeV/$c^{2}$. The superscripts $a$ and $b$ represent the decay modes of $ X\to K^{+}K^{-}\eta'$ and $X\to K_{S}^{0}K_{S}^{0}\eta'$, respectively.
  The uncertainties are statistical only.}

 \label{fitresult2370}
 \begin{tabular}{lcc} \hline\hline
                                            &$\eta'\to\gamma\rho^{0}$           &$\eta'\to\pi^{+}\pi^{-}\eta$ \\ \hline
$M_{X(2370)}$ (MeV/$c^{2}$)                 &\multicolumn{2}{c}{$2341.6\pm 6.5$}     \\
$\Gamma_{X(2370)}$ (MeV)                    &\multicolumn{2}{c}{$\phantom{23}117\pm10$} \\
$N(J/\psi \to \gamma X(2370)^{a})$          &$882\pm112$   & $320\pm40$   \\
$N(J/\psi \to \gamma X(2370)^{b})$          &$174\pm47\phantom{2}$    & \phantom{2}$55\pm15$\\ \hline
$N(J/\psi \to \gamma X(2120)^{a})$          &$<553.5$    & \phantom{2}$<187.3$   \\
$N(J/\psi \to \gamma X(2120)^{b})$          &$<88.7$    & $<30.0$\\ \hline\hline

\end{tabular}
\end{center}
\end{table}

\begin{table}[htp]
\begin{center}
 \caption{ Summary of the MC detection efficiencies
 of the signal yields for the two $\eta'$ modes where the $K\bar{K} \eta'$ invariant-mass
 is constrained to the applied fitting range between 2.0 and 2.7 GeV/$c^{2}$.
 The superscripts $a$ and $b$ represent the decay modes of
 $ X\to K^{+}K^{-}\eta'$ and $X\to K_{S}^{0}K_{S}^{0}\eta'$, respectively.}
 \label{effi_all}
 \begin{tabular}{lcc} \hline\hline
Decay modes           & $\varepsilon_{\eta'\to\gamma\rho^{0}}$    &$\varepsilon_{\eta'\to\pi^{+}\pi^{-}\eta}$ \\ \hline
$J/\psi \to \gamma X(2370)^{a}$            & 12.9 \%            &8.0 \% \\
$J/\psi \to \gamma X(2370)^{b}$            & \phantom{2}8.1 \%  &4.4 \% \\ \hline
$J/\psi \to \gamma X(2120)^{a}$            & 10.3 \%            &6.0 \%  \\
$J/\psi \to \gamma X(2120)^{b}$            & \phantom{2}7.9 \%  &4.6 \%  \\ \hline\hline
\end{tabular}
\end{center}
\end{table}

There is no obvious signal of the $X(2120)$ found in the $K\bar{K}\eta'$ invariant-mass distribution. We performed a simultaneous unbinned maximum-likelihood fit to the $K\bar{K}\eta'$ invariant-mass distribution in the range of [2.0, 2.7] GeV/$c^2$.
The signal, $X(2120)$, is described with an efficiency-weighted BW function convolved with a double Gaussian function.
The mass and width of the BW function are fixed to previously published BESIII results~\cite{PRL1}.
The backgrounds are modeled with the same components as used in the fit of the $X(2370)$ as mentioned above.
The contribution from the $X(2370)$ is included in the fit and its mass, width and the number of events are set free. 
The distribution of normalized likelihood values for a series of input
signal event yields is taken as the probability density function (PDF)
for the expected number of events.  The number of events at 90$\%$ of
the integral of the PDF from zero to the given number of events is
defined as the upper limit, $N^{UL}$, at the 90$\%$ confidence level (C.L.).
We repeated this procedure with different signal shape parameters of $X(2120)$ (by varying the values of mass and width with 1$\sigma$ of the uncertainties cited from~\cite{PRL1}), fit ranges, $\eta'$ sideband regions and background shapes,
and the maximum upper limit among these cases is selected. The statistical significance of the $X(2120)$ is determined to be 2.2$\sigma$.
To calculate  $N^{UL}$ for the $\jp\to\gamma X(2120)\to\gamma\kk\eta'$ ($\jp\to\gamma X(2120)\to\gamma\ksks\eta'$) channel,
the number of signal events for $\jp\to\gamma X(2120)\to\gamma\ksks\eta'$ ($\jp\to\gamma X(2120)\to\gamma\kk\eta'$) channel is left free.
The obtained upper limits of the signal yields are listed in Table~\ref{fitresult2370},
and the upper limit for the product branching fractions are calculated to be
$\mathcal{B}(\jp\to\gamma X(2120)\to\gamma K^{+} K^{-} \eta') < 1.41\times10^{-5}$
and $\mathcal{B}(\jp\to\gamma X(2120)\to\gamma K_{S}^{0} K_{S}^{0} \eta') < 6.15\times10^{-6}$, respectively.

\begin {table*}[htp]
  \begin{center}
	{\caption {Absolute systematic uncertainties of resonance parameters of Mass ($M$, in MeV/$c^{2}$) and Width ($\Gamma$, in MeV) for the $X(2370)$. The items with * are common uncertainties of both  $\eta'$ decay modes.}
		\label{sumsys2370}}

	\begin{tabular}{lcccccccc}  \hline\hline
	
	\multicolumn{1}{c}{\multirow{3}{*}{Source}} &\multicolumn{4}{c}{$J/\psi\to\gamma K^{+}K^{-}\eta'$}  &\multicolumn{4}{c}{$J/\psi\to\gamma \ksks\eta'$}\\
		&\multicolumn{2}{c}{$\gamma\rho^{0}$} &\multicolumn{2}{c}{$\pi^{+}\pi^{-}\eta$}
	&\multicolumn{2}{c}{$\gamma\rho^{0}$} &\multicolumn{2}{c}{$\pi^{+}\pi^{-}\eta$}\\
		                               &$M$ &$\Gamma$      &$M$ &$\Gamma$      &$M$ &$\Gamma$   &$M$ &$\Gamma$ \\ \hline
    Veto $\pi^{0}$                        &0.0  &1  &0.3  &1         &0.2 &1  &0.2 &1 \\
	Veto $\eta$                           &0.2  &1  &--    &--       &0.2 &1  &--  &--  \\
	Fit range                             &0.1  &3  &0.1  &3         &0.1 &3  &0.1 &3 \\
	Sideband region                       &0.1  &2  &0.1  &2         &0.2 &1  &0.1 &1 \\
	Chebychev function                    &0.2  &3  &0.1  &3         &0.2 &1  &0.1 &3 \\
    $J/\psi\to K^{*+}K^{-}\eta'+c.c.$     &0.2  &5  &0.2  &5         &0.2 &5  &0.2 &5 \\
    X(2120)*                               &5.7  &7  &5.7  &7         &5.7 &7  &5.7 &7\\
	Total                                 &5.7  &10 &5.7  &10        &5.7 &9  &5.7 &10\\\hline\hline
\end{tabular}
\end{center}
\end{table*}

\section{SYSTEMATIC UNCERTAINTIES} \label{sec::sys}

Several  sources of systematic uncertainties are considered
for the determination of the mass and width of the $X(2370)$ and the product branching fractions.
 These include the efficiency differences between data and MC simulation in
the MDC tracking, PID, the photon detection, $\ks$ reconstruction, the kinematic fitting, and the mass window requirements of $\pi^{0}$, $\eta$, $\rho$ and $\eta'$.
Furthermore,  uncertainties associated with the fit ranges, the background shapes,
the sideband regions, the signal shape parameters of $X(2120)$, intermediate resonance decay branching
fractions and the total number of $\jpsi$ events
are considered.

\begin{table}[htp]
	\begin{center}
		{\caption {Systematic uncertainties for determination of the branching fraction of $\jp\to\gamma X(2370)\to\gamma K \bar{K} \eta'$ (in \%).
The items with * are common uncertainties of both  $\eta'$ decay modes. I and II represent the decay modes of
 $ \eta'\to\gamma\rho^{0}$, $\rho^{0}\to\pi^{+}\pi^{-}$ and $\eta'\to\pipieta$, $\eta\to\gamma\gamma$, respectively. }
			\label{sumsysBR2370}}
		
		\begin{tabular}{lcccc}  \hline\hline
	\multicolumn{1}{c}{\multirow{2}{*}{Source}} &\multicolumn{2}{c}{$ K^{+}K^{-}\eta'$}&\multicolumn{2}{c}{$ K_{S}^{0}K_{S}^{0}\eta'$} \\
					&I &II &I &II\\ \hline

            MDC tracking*                          &4.0   &4.0   &2.0   &2.0\\
            Photon detection*                      &2.0   &3.0   &2.0   &3.0\\
            $K_{S}^{0}$ reconstruction*            &--    &--    &3.0   &3.0\\
            PID*                                   &4.0   &4.0   &--    &-- \\

			Kinematic fit                          &1.7   &1.0   &3.8   &2.2\\
			$\rho$ mass window                     &0.2   &--    &0.3   &--\\
			$\eta'$ mass window                    &0.1   &0.4   &0.1   &0.3\\	
			Veto $\pi^{0}$                         &1.2   &1.6   &1.7   &0.6 \\
			Veto $\eta$                            &1.0   &--    &0.6   &--\\
			
			Fit range                              &2.4   &2.4   &1.7   &1.7\\
            Sideband region                        &5.4   &2.8   &2.8   &1.2\\
			Chebychev function                     &4.9   &5.5   &1.7   &1.7\\		
            $J/\psi\to K^{*+}K^{-}\eta'+c.c.$       &4.0   &4.0   &2.2   &2.2 \\
             $\mathcal{B}(\eta'\to \gamma\rho^{0}\to\gamma\pippim)$          &1.7   &--    &1.7   &--\\
			$\mathcal{B}(\eta'\to\eta\pippim)$               &--    &1.6   &--    &1.6 \\
			$\mathcal{B}(\eta\to\gamma\gamma)$               &--    &0.5   &--    &0.5\\	
            $\mathcal{B}(K_{S}^{0}\to\pippim)$*              &--    &--    &0.1   &0.1\\
            Number of $\jpsi$ events*              &0.5   &0.5   &0.5   &0.5\\
            Quantum number of $X$                  &16.7  &13.6  &16.0  &19.0\\
            X(2120)*                                &33.7  &33.7  &30.5  &30.5\\
			Total                                  &39.2  &37.7  &35.3  &36.5\\\hline \hline

		\end{tabular}
	\end{center}
\end{table}

\begin{table}[htp]
	\begin{center}
		{\caption {Systematic uncertainties for the determination of the upper limit of the branching fraction of $\jp\to\gamma X(2120)\to\gamma K \bar{K} \eta'$(in \%).
The items with * are common uncertainties of both $\eta'$ decay modes.   I and II represent the decay modes of
 $ \eta'\to\gamma\rho^{0}$, $\rho^{0}\to\pi^{+}\pi^{-}$ and $\eta'\to\pipieta$, $\eta\to\gamma\gamma$, respectively.}
			\label{sumsysBR2120}}
		
		\begin{tabular}{lcccc}  \hline\hline			
			\multicolumn{1}{c}{\multirow{2}{*}{Source}} &\multicolumn{2}{c}{$K^{+}K^{-}\eta'$}&\multicolumn{2}{c}{$K_{S}^{0}K_{S}^{0}\eta'$} \\
			&I &II &I &II\\ \hline
			MDC tracking*                          &4.0   &4.0   &2.0   &2.0\\
            Photon detection*                      &2.0   &3.0   &2.0   &3.0\\
            $K_{S}^{0}$ reconstruction*            &--    &--    &3.0   &3.0\\
            PID*                                   &4.0   &4.0   &--    &-- \\
			
			Kinematic fit                          &1.7 &0.8 &4.0  &3.5\\
			$\rho$ mass window                     &0.2&--&0.3&--\\
			$\eta'$ mass window                    &0.1&0.1&0.2&0.2\\	
			Veto $\pi^{0}$                         &0.8&1.0&1.4 &1.5 \\
			Veto $\eta$                            &0.8&--&1.4&--\\
			
            $\mathcal{B}(\eta'\to \gamma\rho^{0}\to\gamma\pippim)$            &1.7 &-- &1.7  &--\\
			$\mathcal{B}(\eta'\to\eta\pippim)$               &-- &1.6  &--   &1.6 \\
			$\mathcal{B}(\eta\to\gamma\gamma)$               &--   &0.5 &--   &0.5\\	
            $\mathcal{B}(K_{S}^{0}\to\pippim)$*              &--    &--    &0.1  &0.1\\
            Number of $\jpsi$ events*              &0.5  &0.5  &0.5  &0.5\\
            Quantum number of $X$                  &18.2  &16.4  &20.9  &19.3\\
			Total                                  &19.3 &17.6 &21.8 &20.2\\\hline \hline
		\end{tabular}
	\end{center}
\end{table}

\subsection{Efficiency estimation}
The MDC tracking efficiencies of charged pions and kaons are investigated using nearly background-free (clean) control samples of $\jp\to p\bar{p}\pippim$ and $\jp\to\ks K^{\pm}\pi^{\mp}$~\cite{MDCpi, MDC}, respectively. The difference in tracking efficiencies between data and MC is 1.0\% for each charged pion and kaon.
The photon detection efficiency is studied with a clean sample of
$J/\psi\to\rho^{0}\pi^{0}$~\cite{Photon}, and the result shows that
the difference of photon detection efficiencies between data and MC
simulation is 1.0\% for each photon.
The systematic uncertainty from $K_{S}^{0}$ reconstruction is
determined from the control samples of $\jp\to K^{*\pm}K^{\mp}$ and
$\jp\to\phi\ks K^{\pm}\pi^{\mp}$, which indicate that
the efficiency difference between data and MC is less than 1.5$\%$ for each
$\ks$. Therefore, 3.0$\%$ is taken as the systematic uncertainty for the two
$\ks$ in  $\jp\to\gamma\ksks\eta'$ channel.

For the decay channel of $J/\psi\to\gamma K^{+}K^{-}\eta'$, the PID has been used to identify the kaons and pions.
Using a clean sample of $J/\psi\to p\bar{p}\pi^{+}\pi^{-}$, the PID efficiency of $\pi^{+}/\pi^{-}$ has been studied,
which indicates that the $\pi^{+}/\pi^{-}$ PID efficiency for data agrees with MC simulation within 1\%.
The PID efficiency for the kaon is measured with a clean sample of $J/\psi\rightarrow K^+ K^-\eta$.
The difference of the PID efficiency between data and MC is less than 1\% for each kaon. Hence,
In this analysis, four charged tracks are required to be identified as two pions and two kaons,
4\% is taken as the systematic uncertainty associated with the PID.

The systematic uncertainties associated with the kinematic fit are
studied with the track helix parameter correction method, as described
in Ref.~\cite{4cError}. The differences with
respect to those without corrections are taken as systematic
uncertainties.

Due to the difference in the mass resolution between data and MC,
uncertainties related to the $\rho^{0}$ and $\eta'$ mass window requirements are
investigated by smearing the MC simulation to improve the consistency between data and MC simulation. The differences in the detection efficiency before and after smearing are
assigned as  systematic uncertainties for the $\rho^{0}$ and $\eta'$ mass window
requirement.
The uncertainties from the  $\pi^{0}$ and $\eta$ mass-window requirements are estimated by
varying the mass windows of $\pi^{0}$ and $\eta$, and differences in the resulting branching fractions are assigned as the systematic uncertainties of this item.

Furthermore, we considered the effects arising from different quantum numbers of the $X(2120)$ and $X(2370)$. We generated $\jp\rightarrow\gamma X(2120)$ and $\jp\rightarrow\gamma X(2370)$ decays following a $\mathrm{sin}^2\theta_{\gamma}$ angular distribution. The resulting differences in efficiency with respect to the nominal value are taken as systematic uncertainties.

\subsection{Fit to the signal}
To study the uncertainties from the fit range and $\eta'$ sideband region, the fits are repeated
with different fit ranges and sideband regions, the largest differences among these
signal yields are taken as systematic uncertainties, respectively.
To estimate the uncertainties in the description of various background contributions, we performed
alternative fits with third-order Chebychev polynomials modeling the background of the
 $\kk\eta'$ and $\ksks\eta'$ channels.
The maximum differences
in signal yields with respect to the nominal fit are taken as
systematic uncertainties.
The uncertainties from the background of $J/\psi\to K^{*+}K^{-}\eta' + c.c.$
are estimated by absorbing this component
into a Chebychev polynomial function,
and the differences obtained by using the description with or without
the background component of $J/\psi\to K^{*+}K^{-}\eta' + c.c.$ are taken as systematic uncertainties.
The impact of the $X(2120)$ is also considered as a systematic uncertainty in the study of the $X(2370)$.
 The difference between a fit with and without a $X(2120)$ contribution is taken as a systematic uncertainty
 associated to this item.

\subsection{Others}
Since no evident structures are observed in the invariant-mass distributions of $M(K\eta')$, $M(\bar{K}\eta')$ and $M(K\bar{K})$ for the events with a $K\bar{K}\eta'$ invariant mass within the $X(2370)$ mass region (2.2~GeV/$c^{2} < M_{K\bar{K}\eta'} <$ 2.5~GeV/$c^{2}$),
the systematic uncertainties of the reconstruction efficiency due to the possible intermediate
 states on the $K\eta'$, $\bar{K}\eta'$ and $K\bar{K}$ mass spectra are ignored.
The uncertainties on the intermediate decay branching fractions of $\eta'\to\gamma\rho^{0}\to\gamma\pi^{+}\pi^{-}$, $\eta'\to\pi^{+}\pi^{-}\eta$, $\eta\to\gamma\gamma$ and $K_{S}^{0}\to\pi^{+}\pi^{-}$ are taken from the world average values~\cite{pdg}, which are  $1.7\%$, $1.6\%$, $0.5\%$ and $0.1\%$, respectively.
The systematic uncertainty due to the number of $\jp$ events is determined
as 0.5$\%$ according to Ref.~\cite{jpsinumber}.

A summary of all the uncertainties is shown in Table~\ref{sumsys2370}, ~\ref{sumsysBR2370} and \ref{sumsysBR2120}.
The total systematic uncertainties are obtained by adding all
individual uncertainties in quadrature, assuming all sources to be independent.

The $X(2120)$ and $X(2370)$ are studied via $\jpsi\to\gamma \kk\eta'$ and $J/\psi\to\gamma \ksks\eta'$ with two $\eta'$ decay modes, respectively.
The measurements from the two $\eta'$ decay modes are, therefore, combined by considering the difference in  uncertainties of these two measurements.
 The combined systematic uncertainties are calculated with weighted least squares method~\cite{combinepaper} and the results are shown in Table~\ref{combinesystable}.

\begin{table*}[htp]
\begin{center}
 \caption{\label{combinesystable} Combined results of the structure around 2.34 GeV/$c^{2}$, the measured branching fractions and the upper limits.}
 \begin{tabular}{lc} \hline\hline
$M_{X(2370)}$ (MeV/$c^{2}$)                                                             &$2341.6\pm 6.5\text{(stat.)}\pm 5.7\text{(syst.)}$    \\
$\Gamma_{X(2370)}$ (MeV)                                                                &\phantom{2.}$117\pm10\text{(stat.)}\pm 8\text{(syst.)}$  \\
$\mathcal{B}(J/\psi \to \gamma X(2370)\to \gamma K^{+}K^{-}\eta')$  & $(1.79\pm0.23~\text{(stat.)}\pm 0.65~\text{(syst.)})\times10^{-5}$\\
$\mathcal{B}(J/\psi \to \gamma X(2370)\to \gamma K_{S}^{0}K_{S}^{0}\eta')$   & $(1.18\pm0.32~\text{(stat.)}\pm 0.39~\text{(syst.)})\times10^{-5}$\\
$\mathcal{B}(J/\psi \to \gamma X(2120)\to \gamma K^{+}K^{-}\eta')$                      & $<1.49\times10^{-5}$\\
$\mathcal{B}(J/\psi \to \gamma X(2120)\to \gamma K_{S}^{0}K_{S}^{0}\eta')$              & $<6.38\times10^{-6}$\\ \hline\hline

\end{tabular}
\end{center}
\end{table*}

 \section{RESULTS AND SUMMARY}
Using a sample of $1.31\times10^{9} ~J/\psi$ events collected with the BESIII detector,
the decays of $J/\psi\to\gamma \kk\eta'$ and  $J/\psi\to\gamma \ksks\eta'$ are investigated using the two $\eta'$ decay modes,
$\eta'\to\gamma\rho^{0}(\rho^{0}\to\pi^{+}\pi^{-})$ and $\eta'\to\pi^{+}\pi^{-}\eta( \eta\to\gamma\gamma)$.
The $X(2370)$ is observed in the $K\bar{K}\eta'$ invariant-mass distribution with a statistical significance of 8.3$\sigma$.
The mass and width are determined to be\\
\hspace*{0.6cm}{$M_{X(2370)}=2341.6\pm 6.5\text{(stat.)}\pm5.7\text{(syst.)}$ MeV/$c^{2}$},\\
\hspace*{0.6cm}{$\Gamma_{X(2370)}=117\pm10\text{(stat.)}\pm8\text{(syst.)}$ MeV},\\
\noindent which are found to be consistent with those of the $X(2370)$ observed in the previous BESIII results~\cite{PRL1}. The product branching fractions of $\mathcal{B}(\jp\to\gamma X(2370)\to\gamma K^{+} K^{-} \eta')$ and
$\mathcal{B}(\jp\to\gamma X(2370)\to\gamma K_{S}^{0} K_{S}^{0} \eta')$ are measured to be
$(1.79\pm0.23~\text{(stat.)}\pm0.65~\text{(syst.)})\times10^{-5}$ and
$(1.18\pm0.32~\text{(stat.)}\pm0.39~\text{(syst.)})\times10^{-5}$, respectively.
No evident signal for the $X(2120)$ is observed in the $K\bar{K}\eta'$ invariant-mass distribution.
For a conservative estimate of the upper limits of the product branching fractions of
$J/\psi\to\gamma X(2120)\to\kk\eta'$ and $J/\psi\to\gamma X(2120)\to\ksks\eta'$,
the multiplicative uncertainties are considered by convolving the normalized likelihood
function with a Gaussian function.
Upper limits for product branching fractions at 90\% C. L. are determined to be $\mathcal{B}(\jp\to\gamma X(2120)\to\gamma K^{+} K^{-} \eta') < 1.49\times10^{-5}$
and $\mathcal{B}(\jp\to\gamma X(2120)\to\gamma K_{S}^{0} K_{S}^{0} \eta') <  6.38\times10^{-6}$.

To understand the nature of the $X(2120)$ and $X(2370)$, it is critical to measure their spin and parity and to search for them in more decay modes. A partial-wave analysis is needed to measure their masses and widths more precisely, and to determine their spin and parity. This might become possible in the future with the foreseen  higher statistics of $\jp$ data samples.

\begin{acknowledgements}
The BESIII collaboration thanks the staff of BEPCII and the IHEP computing center for their strong support. This work is supported in part by National Key Basic Research Program of China under Contract No. 2015CB856700; National Natural Science Foundation of China (NSFC) under Contracts Nos. 11625523, 11635010, 1332201, 11735014, 11565006; National Natural Science Foundation of China (NSFC) under Contract No. 11835012; the Chinese Academy of Sciences (CAS) Large-Scale Scientific Facility Program; Joint Large-Scale Scientific Facility Funds of the NSFC and CAS under Contracts Nos. U1532257, U1532258, U1732263, U1832207; CAS Key Research Program of Frontier Sciences under Contracts Nos. QYZDJ-SSW-SLH003, QYZDJ-SSW-SLH040; 100 Talents Program of CAS; INPAC and Shanghai Key Laboratory for Particle Physics and Cosmology; German Research Foundation DFG under Contract No. Collaborative Research Center CRC 1044; Istituto Nazionale di Fisica Nucleare, Italy; Koninklijke Nederlandse Akademie van Wetenschappen (KNAW) under Contract No. 530-4CDP03; Ministry of Development of Turkey under Contract No. DPT2006K-120470; National Science and Technology fund; The Knut and Alice Wallenberg Foundation (Sweden) under Contract No. 2016.0157; The Royal Society, UK under Contract No. DH160214; The Swedish Research Council; U. S. Department of Energy under Contracts Nos. DE-FG02-05ER41374, DE-SC-0010118, DE-SC-0012069; University of Groningen (RuG) and the Helmholtzzentrum fuer Schwerionenforschung GmbH (GSI), Darmstadt.
\end{acknowledgements}

\end{document}